\documentclass[pra, superscriptaddress, amsmath, amssymb, aps, twocolumn]{revtex4}
\usepackage[utf8]{inputenc}
\usepackage[english]{babel}
\usepackage{graphics}
\usepackage{graphicx}
\usepackage{float}
 \usepackage{color}
 \usepackage{soul}
 \usepackage[colorlinks=true,linkcolor=blue]{hyperref}
 \usepackage{enumitem}

\usepackage{bm}

\newcommand{\tmu}{\tilde{\mu}}
\newcommand{\tchi}{\tilde{\chi}}
\newcommand{\tpsi}{\tilde{\psi}}
\newcommand{\cN}{\mathcal{N}}
\newcommand{\rev}[1]{\textcolor{black}{{#1}}}

\definecolor{RED}{rgb}{1, 0, 0}

\begin{document}
\allowdisplaybreaks

\title{Formation of nonlinear modes in one-dimensional quasiperiodic lattices with a mobility edge}
\author{Dmitry A. Zezyulin}
\affiliation{School of Physics and Engineering, ITMO University, 197101 St. Petersburg, Russia}
\affiliation{Institute of Mathematics, Ufa Federal Research Center, Ufa 450008, Russia}

\author{Georgy L. Alfimov}
\affiliation{Institute of Mathematics, Ufa Federal Research Center, Ufa 450008, Russia}
\affiliation{Moscow Institute of Electronic Engineering, Zelenograd, Moscow, 124498, Russia}

\begin{abstract}
	We investigate the formation of steady states in    one-dimensional  Bose-Einstein condensates of repulsively interacting ultracold atoms loaded into a quasiperiodic potential  created by two incommensurate periodic lattices. We study the transformations   between linear and nonlinear modes  and describe the  general patterns that govern the birth of  nonlinear modes   emerging in spectral gaps near band edges. We show that nonlinear modes in a symmetric potential undergo   both symmetry-breaking pitchfork bifurcations and saddle-node bifurcations, mimicking the  prototypical behaviors   of symmetric and asymmetric double-well potentials. The properties of the nonlinear modes differ  for bifurcations occurring  below and above the mobility edge. 	In the generic case, when the quasiperiodic potential consists of two incommensurate lattices with a nonzero phase shift between them,  the formation of localized modes in the  spectral gaps occurs through a cascade of saddle-node bifurcations. Because of  the  analogy between the Gross-Pitaevskii equation and the nonlinear Schr\"odinger equation, our results can  also be  applied to optical modes guided by quasiperiodic photonic lattices. 
\end{abstract}
\maketitle

\section{Introduction}

The elevated  interest in quasiperiodic systems originates in  the  intermediate position they occupy between periodically patterned   and fully  disordered media. Quasiperiodic structures can be implemented in various physical environments which, in particular, include atomic Bose-Einstein condensates (BECs) loaded in a superposition of incommensurate optical lattices \cite{Lye,Roati,Reeves,Luschen,Modugno10},  exciton-polariton condensates emerging in etched nanowires \cite{Tanese,Goblot}, and photonic lattices \cite{Lahini,Bellingeri17,Valy13,Arie2010}.   Examples of prominent features that are inherent to disordered systems and have already been experimentally observed using the paradigm of quasiperiodicity include   the Anderson localization  \cite{Modugno10} and the corresponding  phase transition \cite{Lahini}, the existence of a mobility edge \cite{Luschen,Wang2022},  \rev{i.e., the critical energy separating spatially localized and extended states}, and  the fractal energy spectrum \cite{Tanese}. 

The physical systems mentioned above gravitate towards each other  due to the  similarity between the Shr\"odinger equation and the paraxial equation for electromagnetic waves \cite{Segev,Longhi,Carretero}.  Moreover, all these   systems   exhibit significant nonlinear responses and     enable the  existence of solitons and, more generally, steadily propagating  nonlinear  localized waves. Hence the formation of nonlinear patterns in quasiperiodic lattices becomes a relevant topic.  Solitons  emerging in gaps of the  fractal spectrum have been  predicted to exist in  BECs with repulsive interactions \cite{Sakaguchi2006} whose nonlinear effect is akin to the selfdefocusing    in  optical media, where analogous gap solitons have been found too \cite{Huang19}. Solitons  in the semiinfite gap of  quasiperiodic  selffocusing media  were considered in \cite{Kominis2008}. Much attention has been given to nonlinear matter waves \cite{Sakaguchi2006,Baizakov,Burlak12,Burlak07} and optical solitons   \cite{Xie03,Huang16,Huang21,Kartashov21,Kartashov23,Liu2024,Ablowitz06,Ablowitz10,Ablowitz12,Bagchi2012} in  two-dimensional quasiperiodic  and, more generally, quasicrystalline    lattices. Optical solitons in photonic quasicrystals have been produced in several  experiments \cite{Freedman06,Wang2020,Fu2020}. Solitons and solitonlike states in discrete \cite{Johan1,Johan2,Sukho06,Marti10} and layered \cite{Grig10,Huang21OE,Su22,Takahashi12} aperiodic  media have also been studied.

As compared to   periodic systems, which can be exhaustively described with the Floquet-Bloch  theory, the linear spectra of quasiperiodic systems are usually  organized in a much more complex way and contain both localized and spatially extended eigenfunctions. The point spectrum of     localized states has a fractal structure, and     spatially extended states   no longer obey  the Floquet  theorem. The intricate structure of the linear spectrum and the presence of  the mobility edge between localized and extended eigenstates  are expected to   manifest   in the nonlinear regime as well.  While nonlinear states  emerging in quasiperiodic lattices have been documented in numerous studies, the interplay between linear and nonlinear states and the mechanism of  formation of localized nonlinear  modes are not yet fully understood. The main goal of this paper  is to elucidate  the  typical patterns accompanying the formation  of  nonlinear modes in quasiperiodic potentials through  the transformation between linear and nonlinear states below and above the mobility edge.  This issue is  particularly relevant  in atomic BECs, where  the  transition between effectively linear and nonlinear regimes can be controlled    using  the Feshbach resonance management \cite{Kevrekidis2006,BraKon,Deissler10,Wang2022}.

More specifically, we consider a BEC trapped in  a  one-dimensional symmetric bichromatic potential   composed of two incommensurate periodic lattices. The most robust nonlinear states are expected to exist in gaps of the spectrum,  where   the excitation of linear modes is inhibited. We therefore inspect the structure of the spectral band  edges and study families of nonlinear modes that emerge in the  spectral   gaps near the band  edges below and above the mobility edge. For bands situated below the mobility edge, we describe two mechanisms for the formation of nonlinear modes. In the first case, nonlinear modes of infinitesimal amplitude branch off from the  zero solution and form nonlinear families that bifurcate from linear localized states. The second scenario involves the  hybridization  between different   modes  which results in the  formation of new  families through a bifurcation.    We find  that  even the simplest spatially symmetric  bichromatic potential  composed of two  lattices with   zero phase shift  exhibits   remarkably versatile behavior, and its nonlinear modes   undergo   either   pitchfork or saddle-node bifurcations. In a more general case, when the   potential is spatially asymmetric and    composed  of two incommensurate lattices with a  nonzero phase shift, the formation of  localized nonlinear modes    occurs through a cascade of saddle-node bifurcations.   Further, we establish that the existence of the mobility edge dichotomizes  the formation of nonlinear   modes. Below the mobility edge,   a mode is already localized in the linear limit, and hence it remains localized as the nonlinearity increases. Above the mobility edge,  a spatially extended linear mode  builds up its localization  as the chemical potential departs from the band edge and the effective nonlinearity grows. These distinct behaviors lead to  different dependencies of the   number of atoms in the condensate on its chemical potential near the band edge  below and above the mobility edge.  

We approach the   problem using  the so-called approximant path \cite{Zilber2021} which replaces the  quasiperiodic system with infinite spatial extent by  a   finite  system   truncated by periodic boundary conditions. The qualitative patterns described in our study are  robust against    changes in the  accuracy of the rational approximation and highlight   the interrelation   between the fractal linear spectrum and the multitude of emerging nonlinear modes.

The paper is organized as follows. In Sec.~\ref{sec:model} we introduce the model, discuss its rational approximations, and summarize   relevant information about the spectrum of linear states. \rev{The main results of our study are presented in Sections~\ref{sec:edges} and \ref{sec:modes}}.  Section~\ref{sec:edges}    describes the structure of   band edges and discusses their transformation with the change of   accuracy of the rational approximation.  Section~\ref{sec:modes} elucidates the general patterns associated with the formation  of nonlinear modes  near the band edges. This section incorporates   the case of   symmetric quasiperiodic potential and  the more general case of   asymmetric potential created by a pair of incommensurate lattices with  nonzero phase shift. Section~\ref{sec:concl} provides conclusion and outlook.

\section{ {Model, methods,  and  a  summary of results for the linear spectrum} }

\label{sec:model}

We consider a quasi-one-dimensional Bose--Einstein condensate of ultracold atoms and assume that its   meanfield dynamics  obeys   the Gross-Pitaevskii equation:
\begin{equation}
\label{eq:GP_nond}
i \frac{\partial \Psi}{\partial t}= H\Psi+  g |{\Psi}|^2 \Psi, \quad H:=- \frac{1}{2}\frac{\partial^2}{\partial x^2}+V(x).  
\end{equation}
Here $\Psi(x,t)$ is the properly normalized dimensionless order parameter, and $g\geq 0$ represents the corresponding nonlinearity coefficient.  In this study, we will use two different values of the nonlinearity coefficient: the case $g=0$ corresponds to the noninteracting (i.e., linear) regime, and the case $g=1$ corresponds to repulsive interactions. At the same time, the norm of the wavefunction $\cN = \int_{-\infty}^\infty |\Psi|^2dx$  and, respectively, the number of atoms in the condensate remain  free parameters.  Equivalently, we can fix the number of particles  imposing the normalization $\cN=1$ and letting $g$ vary as   a free parameter. These two settings are   interrelated by a simple renormalization.

An essential portion of our results is devoted to a bichromatic  quasiperiodic potential of the form  
\begin{equation}
\label{pot-irrat}
V(x) =v_1 \cos{(2x)}+v_2 \cos{(2\varphi x)},
\end{equation}
where $v_1, v_2 >0$ and    $\varphi$ is an irrational number. Potential (\ref{pot-irrat})   in fact corresponds to a special situation, because (i) it is    symmetric with respect to the origin $x=0$, and (ii)   it achieves   its   maximum (both local and global) at $x=0$. In this respect the bichromatic lattice given by Eq.~(\ref{pot-irrat}) resembles a double-well potential with the barrier situated at $x=0$.  In a more general situation, the bichromatic lattice comprises a combination of mutually shifted incommensurate dependences. This more general case will be    discussed below in Sec.~\ref{sec:shift}. 

To put our work into   proper context, we  notice that the     potential (\ref{pot-irrat})     differs  from the bichromatic  quasiperiodic lattice  previously studied in Ref.~\cite{Sakaguchi2006}. That lattice   had a global  \textit{minimum} at $x=0$ and  no  global maximum.

In the normalized model (\ref{eq:GP_nond})--(\ref{pot-irrat}), the  amplitudes of the lattices, $v_{1,2}$, are measured in units of the recoil energy of one of the lattices. Regarding the normalization of the   order parameter, for $^{87}$Rb atoms with a scattering length   of approximately 100 times the Bohr radius ($a_s \approx 100 a_0$)  and for the lattice  created  by    lasers with wavelengths $\sim 1\,\mu$m and $\sim \varphi\,  \mu$m, respectively, we  can estimate the number of particles in the condensate  as $N \sim 10^4 g  \cN$, where $\cN$ is the `normalized' number of atoms   defined above.

The steady states of the condensate can be represented as $\Psi = e^{-i \mu t} \psi(x)$, where $\mu$ is the chemical potential and $\psi(x)$ is a stationary macroscopic wave function that solves the time-independent Schrödinger equation
\begin{equation}
\label{eq:stat}
\mu \psi = H\psi +  g |{\psi}|^2 \psi.
\end{equation}
We consider real-valued solutions $\psi(x)$, so $|\psi|^2 = \psi^2$.

The main focus   of this  work   is on the   bifurcations between  the eigenmodes of the nonlinear problem (\ref{eq:stat}) and  the  eigenstates of the underlying linear problem (obtained by setting $g=0$)
\begin{equation}
\label{eq:statlin}
\tmu \tpsi = H\tpsi.
\end{equation}
Hereafter we use tildes to distinguish quantities that relate  to the linear regime. 

We treat  the formulated problems  using the so-called  approximant path \cite{Zilber2021}  (see also \cite{Modugno,Diener,LiLiSarma2017,Prates22}) which approximates the    irrational number $\varphi$ by  a rational fraction: $\varphi \approx  p/q$, where $p$ and $q$ are coprime integers. The fraction $p/q$ is called a rational approximation (RA) of the    irrational number $\varphi$. Accordingly, the quasiperiodic potential $V(x)$ given by Eq.~(\ref{pot-irrat}) is replaced by the following expression:
\begin{equation}
\label{eq:V_quasi}
V_q(x):=v_1 \cos(2x)+v_2 \cos\left(2 {p} x/ {q}  \right).
\end{equation}
In this study,  we use the irrational number given by the golden ratio,  $\varphi =(1+\sqrt{5})/2$. Then   a series of optimal approximants of increasing accuracy  is given by   fractions of  subsequent  Fibonacci numbers:
\begin{equation}
\varphi \approx \ldots, \frac{55}{34}, \quad \frac{89}{55}, \quad \frac{144}{89}, \ldots
\end{equation}
The greater the denominator, the more accurate the approximation. 

In contrast to  the quasiperiodic potential $V(x)$,  the   potential $V_q(x)$  introduced in Eq.~(\ref{eq:V_quasi}) is   \textit{periodic}. Its   period   is equal to $L_q :=\pi q$. The difference between $V_q(x)$ and $V(x)$ remains small  only  in a bounded  spatial domain. As a result, it becomes relevant to restrict our  consideration to a finite domain by imposing the  periodic boundary conditions. Specifically, we require 
\begin{equation}
\label{bound}
\Psi(-\pi q/2,t)=\Psi(\pi q/2,t)\quad    [\psi(-\pi q/2)=\psi(\pi q/2)].
\end{equation}
The choice of periodic boundary conditions can be  justified    by  several additional observations: (i) For eigenfunctions    that are well localized within the unit cell $I_q := [-\pi q/2, \pi q/2)$ the difference between periodic and zero boundary conditions is  insignificant; (ii)  Linear eigenfunctions  $\tpsi(x)$ satisfying the  boundary conditions (\ref{bound})  actually represent  Bloch functions of the band structure associated with the   periodic potential $V_q(x)$. This fact may be  useful for   for interpreting the results;  (iii)  Periodic boundary conditions are ideal   for efficient numerical methods used to solve   linear eigenproblems  of the form  (\ref{eq:statlin}). To solve  the linear eigenproblem (\ref{eq:statlin}) with the periodic potential $V_q(x)$,  we used the    Fourier collocation method which expands the potential $V_q(x)$ and  the unknown eigenfunctions $\tpsi(x)$ into Fourier series and approximates  the second spatial derivative $\partial_x^2$ in reciprocal space (see e.g. \cite[Ch.~7]{Yang}). By  truncating the Fourier series,  we   reduce the problem  to    evaluating  the spectrum of a certain matrix, whose eigenvalues provide the chemical potentials  $\tmu$ of the BEC in the linear   limit,  and the eigenvectors  are composed  of    Fourier coefficients of the corresponding   eigenfunctions $\tpsi$. For all the RAs used in this work, 4500 Fourier harmonics were sufficient  to ensure  that  the results were numerically accurate,  i.e., the amplitudes of all neglected Fourier  coefficients were comparable to the machine precision. Regarding the nonlinear problem (\ref{eq:stat}), it was solved using Newton's method with the periodic boundary conditions. 

For each computed nonlinear mode, we have also checked its spectral stability against small perturbations. According to the standard procedure  (see e.g. \cite{Yang}), it is obtained from the numerical solution of  the  eigenvalue problem
\begin{eqnarray}
\lambda \eta_1 = (H - \mu + 3 g \psi^2)\eta_2,\,\, \lambda \eta_2 = -(H - \mu +  g \psi^2)\eta_1,
\end{eqnarray}
where $\eta_1(x)$ and $\eta_2(x)$ represent  the  real and imaginary parts of the small perturbation,  respectively, and  $\lambda$ is and eigenvalue that  determines the dynamical behavior of the perturbation   ($\propto e^{\lambda t}$). A growing (or  unstable) perturbation  is associated with  an eigenvalue with a positive real part.

\begin{figure}
	\begin{center}
		\includegraphics[width=0.999\columnwidth]{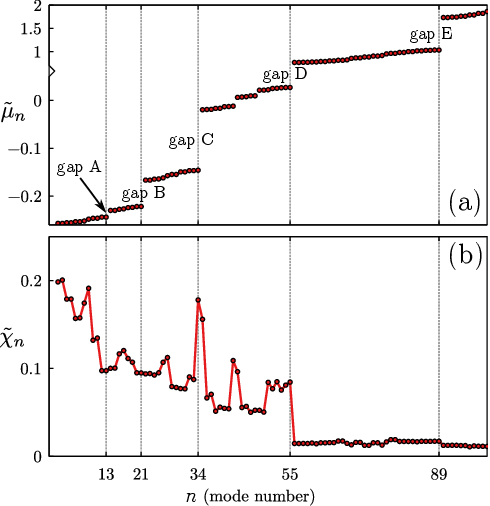}
	\end{center}
	\caption{(a)  Chemical potentials   $\tmu_1<\tmu_2<\ldots\tmu_n<\ldots $ and (b)  IPRs $\tchi_n$ for the 100 lowest linear modes, computed from   the  linear eigenvalue problem (\ref{eq:statlin}) with a rational approximation   $p/q = 89/55$, and amplitudes of the lattices  $v_1 = v_2 = 0.8$. Several gaps are labelled with letters A-E in the upper panel.  The gaps are located at mode numbers $n$ corresponding to the Fibonacci sequence, as highlighted by dotted vertical lines.  The mobility edge is located at $n=55$, so the bands below gap D consist of localised eigenfunctions, and the bands above gap D consist of spatially extended eigenfunctions. We note that the vertical axis is broken in the upper panel.}
	\label{fig:gaps}
\end{figure}

A representative example of the linear eigenspectrum calculated for a specific RA is shown in Fig.~\ref{fig:gaps}. Here, we summarize several facts about the linear spectrum that are relevant to our study. Most of these facts have been discussed in previous studies on bichromatic quasiperiodic potentials (see \cite{Modugno, Biddle, BidSar, Adhikari2009, Yao2019, Prates22, LiLiSarma2017, Konotop2024, Boers2007, Roux2008, Larcher2009, Cheng14, Zhou13, Li16}).  
\begin{enumerate}[label=(\roman*)]	
	\item   The spectrum of the linear eigenvalue problem (\ref{eq:statlin}) is bounded from below and discrete, due to the periodic boundary conditions. We enumerate the chemical potentials in ascending order: $\tmu_1 < \tmu_2 < \ldots < \tmu_n < \ldots$. This discrete spectrum consists of clusters of closely spaced eigenvalues   (which we call \textit{bands} in what follows) separated by \textit{gaps}. It is essential that   gaps are robust with respect to the choice of the RA.  This means that if we choose $p$ and $q$ to be large enough, any further increase in their values will lead to an increase in the number of eigenvalues within the bands (which become more densely populated), while the locations and widths of the gaps remain relatively unchanged.
	
	\item For $\varphi$ equal to the golden ratio, the locations of the gaps in the discrete spectrum correspond to the Fibonacci numbers. Therefore, the gaps are located between $\tmu_{13}$ and $\tmu_{14}$, $\tmu_{21}$ and $\tmu_{22}$, $\tmu_{34}$ and $\mu_{35}$, etc. (see Figure~\ref{fig:gaps} for an illustration).

	\item For sufficiently large  values of $v_1$ and $v_2$,  the spectrum  exhibits a  \textit{mobility edge}, i.e., several lower bands consist of localized eigenfunctions, while the the remaining part of the spectrum consists of eigenfunctions that are  extended over the entire period $I_q  = [-\pi q/2, \pi q/2)$.  The localization is quantified  by  the inverse participation ratio  \rev{(IPR)}, given by
	\begin{equation}
	\tchi_n=  \frac {\int_{I_q}   |\tpsi_n(x)|^4 d x} {\left(\int_{I_q} |\tpsi_n(x)|^2 d x\right)^2}.
	\end{equation}
	States  with density extended over the entire   cell $[-\pi q/2, \pi q/2)$  correspond to relatively small IPR values, $\tchi_n \sim 1/(\pi q)$, while localized states have $\tchi_n \gg 1/(\pi q)$.
	
	\item Since the introduced potential \rev{resembles a symmetric double well}  (in the sense explained above), the spectrum of   localized states $\{\tmu_n,   \tpsi_n(x)\}$ below the mobility edge contains multiple   pairs of closely spaced eigenvalues associated with eigenfunctions  that have opposite parities, with  one    eigenfunction being an even (symmetric) function  of $x$: $\tpsi(x) = \tpsi(-x)$, and another eigenfunction being odd (antisymmetric) function: $\tpsi(x) = -\tpsi(-x)$. \rev{If the bichromatic potential has a nonzero phase shift between the two sublattices, then, in the generic case,  the potential is free from any special symmetry. In this   case the eigenfunctions $\tpsi(x)$ are neither even nor odd, and the  pairs of closely spaced eigenvalues are not present in the spectrum.   In this sense, the distribution of eigenvalues becomes `more uniform' within each band. In the meantime, the locations of the band edges do not significantly depend on the value of the phase shift.}

\end{enumerate}

\section{Structure of the  band edges}
\label{sec:edges}

We are interested in the nonlinear states that arise near the edges of linear bands and extend into spectral gaps. As our system is dominated by a repulsive nonlinearity, for which chemical potentials increase with the growth of   condensate density, we investigate bifurcations between linear and nonlinear states that occur at the right edges of bands, i.e., bifurcations from the rightmost eigenvalues in each band. Therefore, it becomes important to take a closer look at the structure of the linear spectrum near band edges. Specifically, we have chosen several gaps labeled A-E in Figure~\ref{fig:gaps}, where bands adjacent to gaps A and D are located below the mobility edge and consist of localized eigenfunctions, while gap E is located above the mobility edge and the adjacent bands contain spatially extended eigenfunctions.

\begin{figure}
	\begin{center}
		\includegraphics[width=0.999\columnwidth]{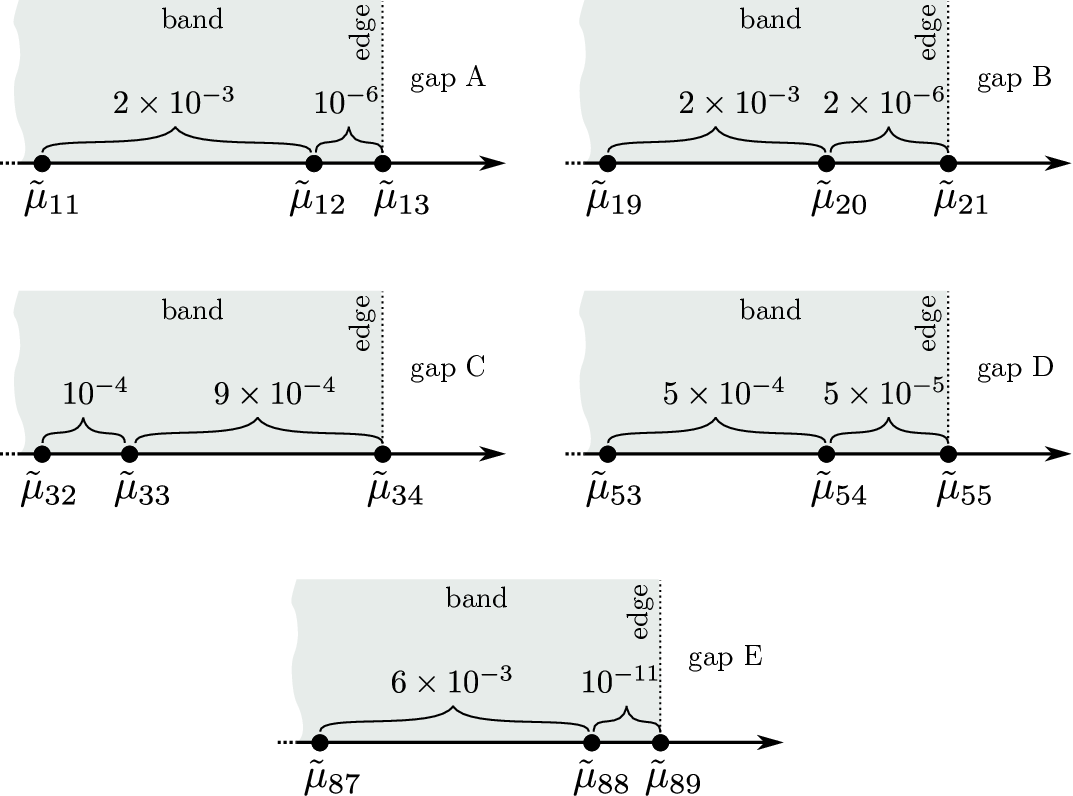}
	\end{center}
	\caption{Schematic illustrations for structure of right band edges adjacent to gaps A-E  from Fig.~\ref{fig:gaps}. Each diagram shows three rightmost chemical potentials $\tmu_{n}$ and indicates distances between them with curly overbraces (not to scale). Shown schematics correspond to  $p/q = 89/55$ and $v_1 = v_2 = 0.8$.  Dotted vertical lines     indicate   the right band edges.	}
	\label{fig:scheme}
\end{figure}

\begin{table*}
	\begin{tabular}{p{4.5cm}|p{2.5cm}p{2.5cm}p{2.5cm}p{2.5cm}}
		&$p/q =  55/34$ & $p/q = 89/55$ & $p/q = 144/89$ & $p/q = 233/144$\\[2mm]\hline
		gap A: $-0.24\lesssim \mu \lesssim -0.23$\qquad\qquad   & s-a/s pair &  s-a/s pair & boundary & new s-a/s pair\\[3mm] 
		gap B: $-0.22\lesssim \mu \lesssim -0.17$ \qquad\qquad &  boundary &  s-a/s pair & s-a/s pair & boundary\\[3mm] 
		gap C: $-0.15\lesssim \mu \lesssim -0.02$ \qquad\qquad &  \multicolumn{4}{c}{   symmetric mode for any $p/q$}\\[3mm] 
		gap D: $-0.15\lesssim \mu \lesssim -0.02$ \qquad\qquad  & s-a/s pair &  s-a/s pair & boundary & new s-a/s pair
	\end{tabular}
	\caption{The structure of the right band edges adjacent to gaps   A-D from Fig.~\ref{fig:gaps}, for different RAs.}
	\label{tbl:gaps}
\end{table*}

\begin{figure}
	\begin{center}
		\includegraphics[width=0.999\columnwidth]{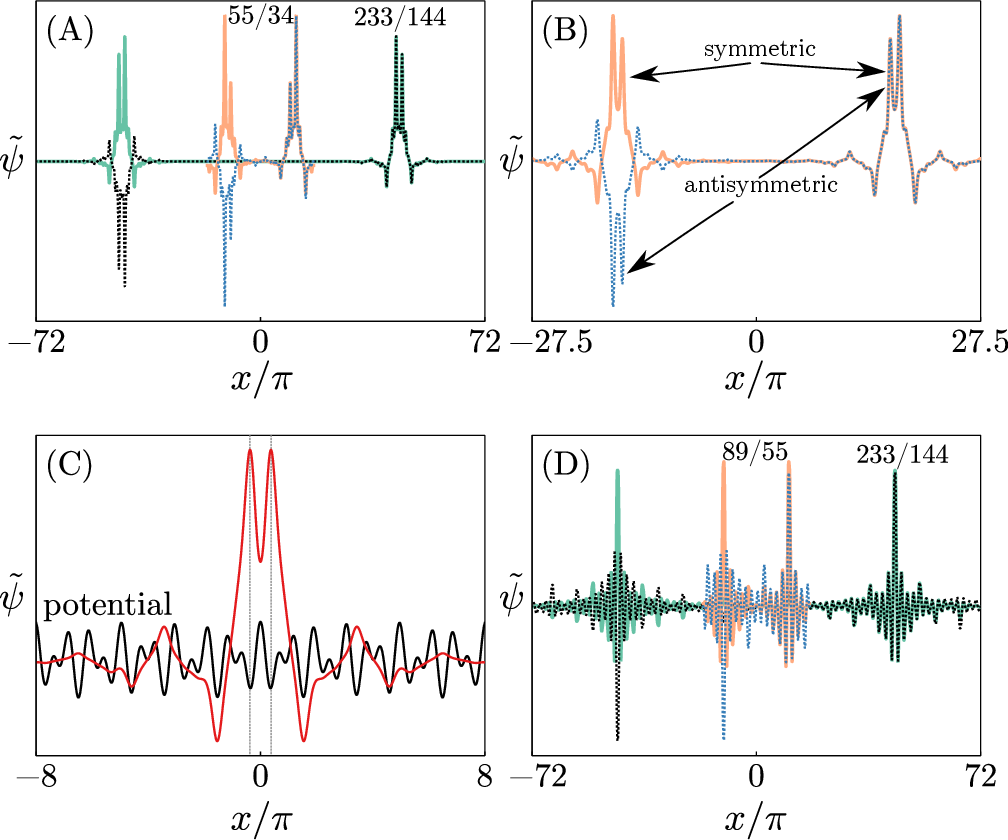}
	\end{center}
	\caption{Eigenfunctions at the right edges of bands   adjacent to gaps A--D  from Fig.~\ref{fig:gaps}. For gap~A, the edge   corresponds to   a  pair of symmetric and antisymmetric eigenfunctions (`s-a/s pair') at $p/q = 55/34$ and to a new s-a/s pair for $p/q = 233/144$ (see the corresponding row in Table~\ref{tbl:gaps}); for gap~B the  band edge corresponds to  the same s-a/s pair for $p/q = 89/55$ and $p/q=144/89$; for gap~C the  right band edge always corresponds to the symmetric  function which is localized around   the origin; for gap~D the right  band edge corresponds to different s-a/s pairs at $p/q = 89/55$ and $p/q = 233/144$. In (C) we additionally plot potential $V_q(x)$.
	}
	\label{fig:eigenf}
\end{figure}

We consider several rightmost eigenstates at the band edge   adjacent to each chosen gap. The results can be summarized  in several  distinctive behaviors summarized in Fig.~\ref{fig:scheme} for   fixed RA with $p/q = 89/55$ and in Table~\ref{tbl:gaps} for several different   RAs of increasing accuracy. In Fig.~\ref{fig:eigenf} we  plot representative examples of eigenfunctions corresponding to  the band edges  for different RAs.

In the first case, a band edge corresponds to a pair of closely spaced eigenvalues associated with localized eigenfunctions of opposite parity. One of the eigenfunctions is symmetric, and the other is antisymmetric.   `Closely spaced'  means that the difference between the   eigenvalues  $\tmu_{n}$ and $\tmu_{n-1}$  is at least an order of magnitude smaller than the difference to the next neighboring eigenstate $\tmu_{n-2}$.  For  example,  $|\tmu_{13} - \tmu_{12}| \ll |\tmu_{12} - \tmu_{11}|$  and   $|\tmu_{21} - \tmu_{20}| \ll |\tmu_{20} - \tmu_{19}|$ in   Fig.~\ref{fig:scheme}.  In Table~\ref{tbl:gaps} this situation is labeled with writings `s-a/s pair'. The existence of these  `s-a/s pairs' is a consequence of the similarity between the quasiperiodic lattice and the double-well potential. \rev{Let us now consider a situation when we switch from a less accurate RA to a more accurate one: say, from $p/q=55/34$ to $p/q = 89/55$. Respectively, we widen the domain where the problem is considered:   $ I_{34}\subset I_{55}$, and new localized linear modes appear in the spectrum. As mentioned  above [see point~(i) in Sec.~\ref{sec:model}], the bands are robust with respect to the change of the RA, i.e.,  the locations of  the edges of each band do not change significantly when we switch from one RA to another. However, it is  possible  that  under the more accurate RA   a new pair of symmetric-antisymmetric pair will appear  to the right of  the  rightmost eigenvalue  of the previous (i.e., less accurate) RA. In that case, the right edge of the band   will  change from the old s-a/s pair to the new pair.} In Table~\ref{tbl:gaps}  this situation corresponds to cells labeled  with  `new s-a/s pair'.  Changes in the right band edge from one pair of eigenfunctions to a different pair can be seen   in Fig.~\ref{fig:eigenf}(A) and~(D).

The second common situation  corresponds to the band edge represented by   an eigenfunction that is sharply localized at the boundaries of the selected domain $I_q = [-\pi q/2, \pi q/2)$. The corresponding cells in  Table~\ref{tbl:gaps} are labelled with writings `boundary'. 

The third situation occurs only for the gap C. For any RA, the corresponding band edge has a well-separated, symmetric mode that is localized around the origin. This mode, plotted in Fig.~\ref{fig:eigenf}(C), has two distinctive peaks that coincide with local minima of the potential near the origin, as highlighted by dotted vertical lines in Fig.~\ref{fig:eigenf}(C). This mode is remarkably robust with respect to changes in the RA, due to its strong localization around the origin. Therefore, a transition from a narrower spatial window to a broader one does not affect the shape of the localized eigenfunction. The existence of this type of mode can be expected due to the bicentric structure of the chosen potential. However, its robustness and the fact that it always sits at the right edge of the band are not evident beforehand.

Finally, for  the gap~E, which is located above the mobility edge, we found that, for each considered  RA, except for 144/89, the  right edge of  the corresponding  band is formed  by  a pair of virtually coinciding and spatially extended eigenfunctions (see the corresponding schematics in Fig.~\ref{fig:scheme}E); for instance, for $p/q = 55/34$  we compute $\tmu_{55} - \tmu_{54} \sim 10^{-8}$ and   for $p/q = 89/55$, we found $\tmu_{89} - \tmu_{88} \sim 10^{-11}$. In fact, these   tiny  distances  are comparable to the numerical error.    
For the  RA with    144/89, the band edge corresponds to a single extended eigenfunction which is well separated from a pair of virtually coinciding eigenstates. 

\section{Nonlinear modes near band edges}
\label{sec:modes}


\subsection{General considerations}

Different structures of the  band edges classified in the previous section   lead to different behaviors of    nonlinear modes theta bifurcate   near those bands.  The standard perturbation theory, similar to that developed in \cite{Zez08}, shows that a nonlinear mode $\psi(x)$ emerging   from an isolated linear eigenfunction $\tpsi_n(x)$  obeys the following approximation near   the bifurcation point:
\begin{equation}
\psi(x) \approx \left( \frac{\mu - \tmu_n}{\tchi_n} \right)^{1/2} \tpsi_n(x)  \quad \mbox{for } |\mu - \tmu_n|\ll 1.
\end{equation}
Using the latter approximation, we can calculate the dependence of the number of particles on the chemical potential. We find that, in the leading order, this relationship is a straight line, and its slope is \rev{inversely} proportional to the IPR of the underlying linear mode:
\begin{equation}
\label{eq:slope}
\cN(\mu) \approx  (\mu - \tmu_n) / \tchi_n.
\end{equation}
Hence bifurcations from   more   localized linear modes correspond to relatively small slopes $d\cN(\tmu_n) /d\mu$.   However for bifurcations that occur   above the mobility edge, where  the  localization of linear modes is weak, the  dependence $\cN(\mu)$ is expected to be distinctively different   compared to bifurcations below the mobility edge.

\begin{figure}
	\begin{center}
		\includegraphics[width=0.999\columnwidth]{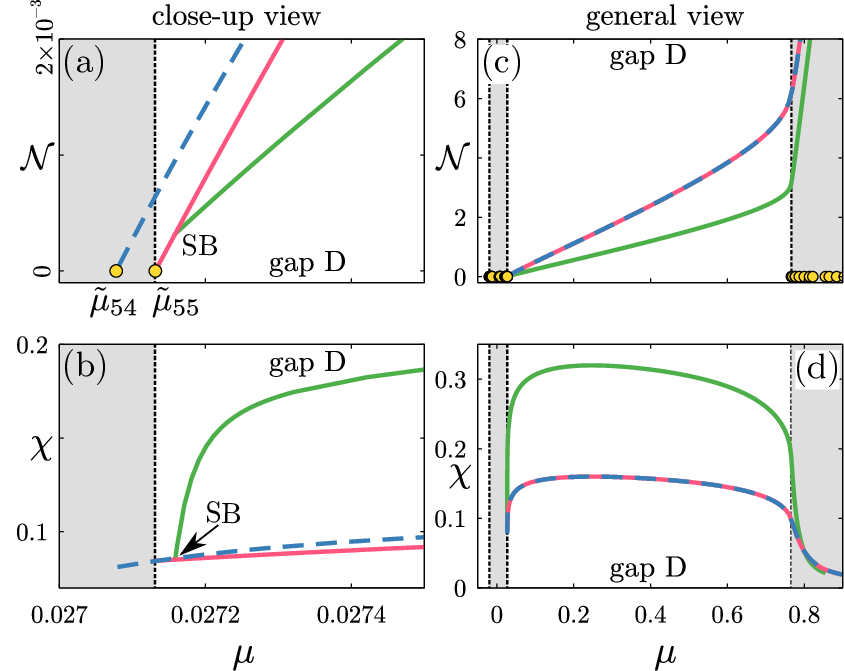}
	\end{center}
	\caption{The number of particles $\cN$ and IPR $\chi$ are plotted against the  chemical potential $\mu$ for the nonlinear modes that  emerge in gap~D from linear modes with numbers $n=54$ and $n=55$. Panels (a,b) show a closer view of the  band edge where the nonlinear families are born, while   panels (c,d) provide a  small-scale  view of   the same dependencies. The shaded areas represent the bands, and the yellow circles in the upper panels indicate positions of the chemical potentials in the linear limit $\tmu_n$. The labels `SB' in (a,b) indicate  the symmetry-breaking pitchfork bifurcation.  In this figure, we use   RA with  to $p/q = 89/55$,  and the lattice amplitudes are set to $v_1 = v_2 = 0.8$.
	}
	\label{fig:gapD}
\end{figure}

\subsection{Pitchfork bifurcation}

From now on, we will fix the parameters   as $p/q= 89/55$ and   $v_1 = v_2 = 0.8$, and we will compute the nonlinear modes that appear  in the gaps  C,  D, and  E which correspond to three different   structures of the band edges.   

We start with families emerging in gap~D, as   the corresponding band  edge  features the most frequent situation   being formed by a pair of symmetric and antisymmetric eigenfunctions with chemical potentials $\tmu_{54}$ and $\tmu_{55}$,
see the corresponding schematics in Fig.~\ref{fig:scheme}D.  The close-up view    in   Fig.~\ref{fig:gapD}(a,b)  shows that both symmetric and antisymmetric eigenstates birth nonlinear families that branch off from the zero solution at $\mu = \tmu_{54}$ and $\mu = \tmu_{55}$, respectively. Additionally,   a symmetry-breaking bifurcation occurs at the antisymmetric family. This   bifurcation is of pitchfork type, as it originates two new families of asymmetric solutions (i.e., not symmetric neither antisymmetric). However, since the emerging asymmetric solutions $\psi_{1,2}$ are interrelated by the parity reversal $\psi_{1}(x) = \psi_{2}(-x)$, they share equal numbers of particles $\cN$ and IPRs $\chi$. As a result, only one bifurcating  curve is visible in Fig.~\ref{fig:gapD}(a,b).

Using the two-mode approach, which assumes that the full nonlinear dynamics can be satisfactorily described in terms of only two modes, with $n=54$ and $n=55$,   the number of particles at which the symmetry-breaking bifurcation occurs can be estimated as \cite{Prates22}
\begin{equation}
\label{eq:Nsb}
\cN_{SB} \approx  {(\tmu_{55} - \tmu_{54})} / {\tchi_{55}},
\end{equation}
where
$\tchi_{55}$ is the   IPR of the corresponding linear mode (since the density distributions  of symmmetric and antisymmetric partners are similar, one can assume that  $\tchi_{55} \approx  \tchi_{54}$). For the numerical values used in Fig.~\ref{fig:gapD}, the approximation (\ref{eq:Nsb}) gives $\cN_{SB} \approx  6\times 10^{-4}$, which is in reasonable agreement  with   the  numerical  value $\cN_{SB} \approx 3\times 10^{-4}$. The discrepancy arises from the limited applicability of the two-mode reduction, which requires  taking into account  more comprehensive multimode dynamics \cite{Prates22}.

The bifurcation diagram in Fig.~\ref{fig:gapD}(a,b) is similar to the symmetry-breaking pitchfork bifurcations that occur in symmetric double-well potentials \cite{TheKev,SBB1,SBB2,SBB3,SBB5,SBB6,SBB7,SBB8,SBB9}. In terms of spectral stability, the   picture is also similar to that  in  symmetric double wells: after the bifurcation, the linearization spectrum of solutions at the antisymmetric family   gets a new  pair of  purely real eigenvalues  of opposite sign,  $\pm \lambda$, and hence the positive eigenvalue implies the instability. The bifurcated asymmetric family is mostly stable  until it reaches the next band, Some fragments of this family and other families considered herein experience  subtle oscillatory instabilities, which   are known  for gap solitons in purely periodic media \cite{Yang,instability}. 
 
For completeness, we mention that in a  recent study in Ref.~\cite{Prates22},   similar symmetry-breaking bifurcations were found to occur  for  antisymmetric states situated near the \textit{left} edge of a band.   As a result of the repulsive nonlinearity, the bifurcated states lose their localization shortly after their formation due to interaction with other modes of the band.  In our case, however, the situation is different, as the bifurcations occur  near the \textit{right} band edge, and the bifurcating  families continue over the entire gap~D preserving their localization, as becomes evident from Figs.~\ref{fig:gapD}(c,d) which provide a   general view of the nonlinear modes in the gap. In fact, as can be seen in Fig.~\ref{fig:gapD}(d), the most localized modes are situated approximately in the center of the gap. When the nonlinear families reach the next band, their IPRs decrease sharply due to the excitation of linear modes.

For gaps~A and B the bifurcation scenario is similar to that of gap~D, because for the chosen RA $p/q = 89/55$  their   band edges have  a similar  structure, as shown in the corresponding column of Table~\ref{tbl:gaps}.

\subsection{Saddle-node bifurcations}

\begin{figure}
	\begin{center}
		\includegraphics[width=0.999\columnwidth]{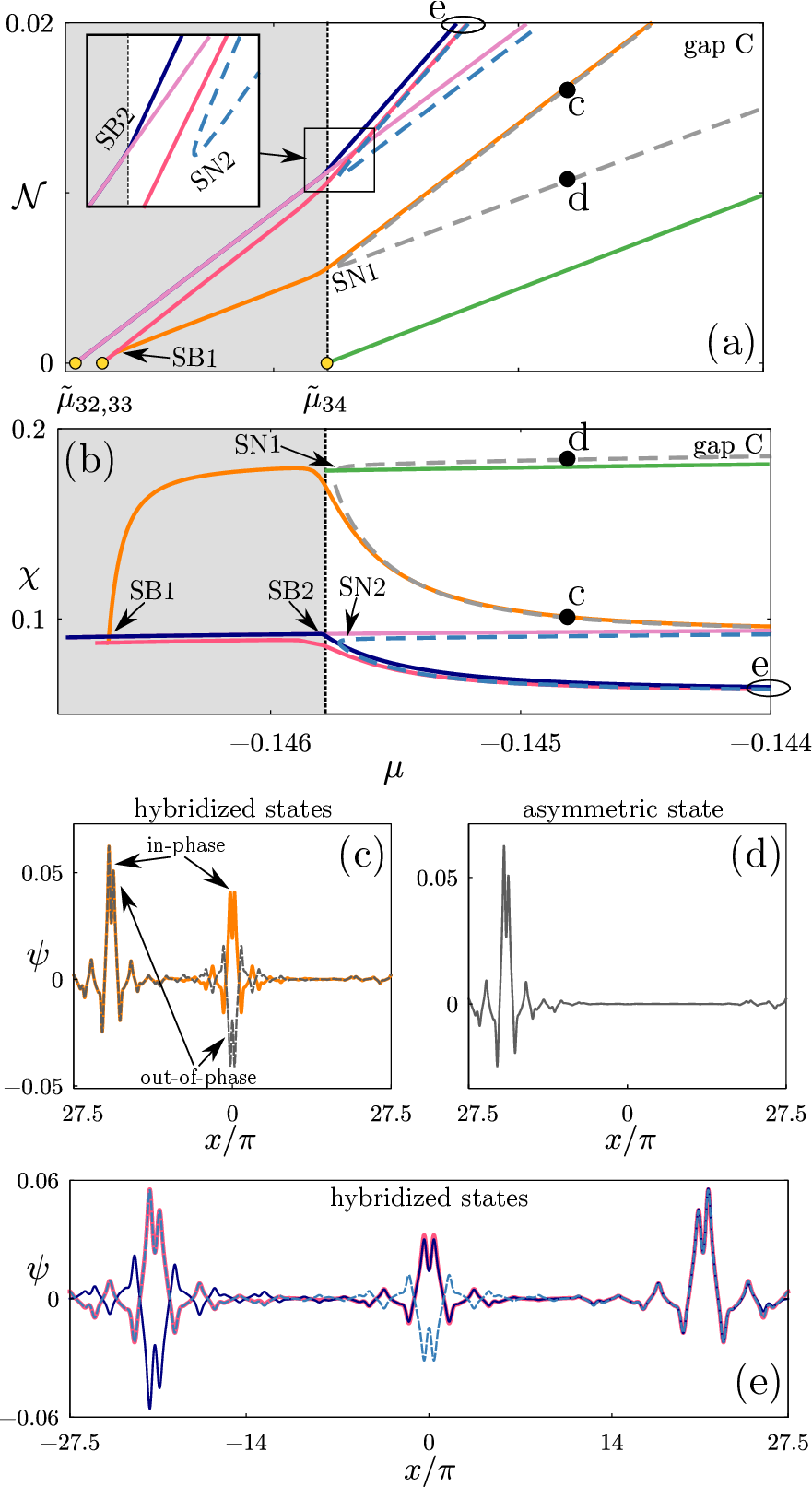}
	\end{center}
	\caption{The number of particles $\cN$ (a) and the IPR $\chi$ (b) vs.  $\mu$ for the nonlinear families  in gap~C. The yellow circles in (a) indicate the chemical potentials $\tmu_n$ of  the   noninteracting BEC.  Labels `SB1,2' and `SN1,2'    indicate   two symmetry-breaking and two   saddle-node bifurcations, respectively. The point `c' in (a,b) corresponds to the in-phase and out-of-phase hybridized states shown in (c), and point `d' corresponds to the asymmetric state shown in (d). Three  more complex  hybridized states with different relative  phases  are shown in (e).  In this figure,  $p/q = 89/55$  and $v_1 = v_2 = 0.8$.
	}
	\label{fig:gapCcl}
\end{figure}

Next, we proceed to gap~C, which is  adjacent to the band edge   formed by   a  separate    single eigenvalue $\tmu_{34}$, while  two preceding  eigenfunctions  with $n=33, 32$  form a symmetric-antisymmetric pair   with closely situated chemical potentials $\tmu_{33}$ and $\tmu_{32}$, as shown  in the   corresponding schematics in Fig.~\ref{fig:scheme}C.  A close-up view of   nonlinear families emerging near this band edge is shown in Fig.~\ref{fig:gapCcl}(a,b). First, we observe that a single family bifurcates from the rightmost eigenvalue $\tmu_{34}$; the shape of   nonlinear solutions in this family is similar to the underlying symmetric linear mode  shown in Fig.~\ref{fig:eigenf}(C). These nonlinear solutions  remain localized with the approximately constant IPR. Regarding the symmetric-antisymmetric pair formed by the linear  eigenstates with chemical potentials  $\tmu_{32}$ and $\tmu_{33}$, we observe a pitchfork symmetry-breaking bifurcation [labelled as `SB1' in Fig.~\ref{fig:gapCcl}(a,b)] with the formation of  a new asymmetric family. A representative asymmetric state is displayed in Fig.~\ref{fig:gapCcl}(d).  As this  asymmetric  family approaches the band edge, it couples to   the rightmost linear state  $\tmu_{34}$, which results in the formation of an in-phase  nonlinear bound state whose profile   is presented in Fig.~\ref{fig:gapCcl}(c). The formation of this bound   state is possible because the two interacting states are localized at different spatial positions.   As a result, the bound state emerges as a  `weakly nonlinear' superposition of the asymmetric state emerging after the SB1-bifurcation and the linear state $\tmu_{34}$ corresponding to the band edge. The formation of this hybridized bound state is accompanied by an abrupt increase in the slope  of  the $\cN(\mu)$ curve in     Fig.~\ref{fig:gapCcl}(a),  while the corresponding IPR curve, $\chi(\mu)$, goes down in Fig.~\ref{fig:gapCcl}(b). This is  because the  IPR of the hybridized   state is lower than the IPRs calculated separately for its constituents. 

At the same time,  the formation of the nonlinear hybridized state is accompanied by the emergence of  a  pair of new families   through  a saddle-node bifurcation labelled  with `SN1' in Fig.~\ref{fig:gapCcl}(a,b). One of these families (with a lower number of particles, $\cN$, and  higher values of IPR, $\chi$) is a virtual  continuation of the asymmetric family. Its representative profile is shown in Fig.~\ref{fig:gapCcl}(d).   The second  family emerging at the SN1 bifurcation (specifically, the family  with larger   $\cN$ and lower $\chi$) consists of out-of-phase   hybridized states exemplified in Fig.~\ref{fig:gapCcl}(c). In-phase and out-of-phase bound states  have nearly identical  density distributions $|\psi(x)|^2$ at the same  chemical potential $\mu$, so their numbers of atoms, $\cN$, and IPRs, $\chi$,   almost    coincide in Fig.~\ref{fig:gapCcl}(a,b).

The hybridization between different   modes is  a common  pattern  responsible for the  formation of a multitude of new  nonlinear families  through  saddle-node and   pitchfork (i.e., symmetry-breaking) bifurcations.   Indeed, in  Fig.~\ref{fig:gapCcl}(a,b) we additionally observe that  coupling between the  family bifurcating from   $\tmu_{33}$ and the linear mode $\tmu_{34}$ results in another saddle-node bifurcation labelled as `SN2', and coupling between  the nonlinear family emerging from   $\tmu_{33}$ and linear mode $\tmu_{34}$ leads to a pitchfork bifurcation which is labelled with `SB2' in Fig.~\ref{fig:gapCcl}(a,b), because this  bifurcation is essentially of the symmetry-breaking type. At this bifurcation, a pair of new   families   branch off with  different relative phases between the hybridized eigenstates, see representative examples in Fig.~\ref{fig:gapCcl}(e).  The emerging nonlinear modes encompass all possible relative phase combinations between the    hybridized wavepackets, which indicates that the bifurcation diagrams presented in Fig.~\ref{fig:gapCcl}(a,b) provide  the complete information about the modes with $n=32,33, 34$.  

Pitchfork and saddle-node bifurcations are typical to occur in symmetric and asymmetric double-well potentials, respectively \cite{TheKev}.   Although  in our case, the potential  given by Eq.~(\ref{pot-irrat}) is a symmetric   function of the spatial coordinate $x$,  saddle-node bifurcations can still occur through the coupling   between localized modes with essentially different shapes, i.e., between modes that  are spatially separated and  do not form a pair of symmetric and antisymmetric states.   Saddle-node bifurcations  are also typical to occur in multiwell potentials \cite{Kapitula2006,Sacchetti2012,Goodman2017}, and in purely periodic potentials,  where more complex solitons can be systematically constructed from   simpler ones using  the `composition relation' \cite{composition} (see also \cite{Akylas2012}). For  quasiperiodic lattices,   saddle-node bifurcations   were recently predicted to exist in Ref.~\cite{Konotop2024} on the basis of a few-mode analysis.


Regarding the stability of the solutions emerging at the  bifurcation SN1, the subfamily with a lower number of atoms $\cN$ contains stable solutions, For the solutions at the upper subfamily, the linear stability spectrum contains a  pair of purely real eigenvalues, which indicates that those solutions are   unstable. For  bifurcation~SN2, the   subfamily with the lower $\cN$ is already unstable, as  it is the virtual continuation of the   symmetric family which loses stability after the pitchfork bifurcation SB1. The upper family is even `more unstable', as its linearization spectrum acquires   an additional pair of real eigenvalues. This difference in  stability properties between the lower and upper subfamilies  is also consistent with the stability patterns  in asymmetric double wells \cite{TheKev}.


As the chemical potential $\mu$ increases, all the families shown in Fig.~\ref{fig:gapCcl}(a,b) continue across gap~C and reach the next  band, where the solutions become delocalized and unstable. This behavior is similar to what was already plotted for gap~D in Fig.~\ref{fig:gapD}, and   therefore it is not shown in Fig.~\ref{fig:gapCcl}.

\begin{figure}
	\begin{center}
		\includegraphics[width=0.999\columnwidth]{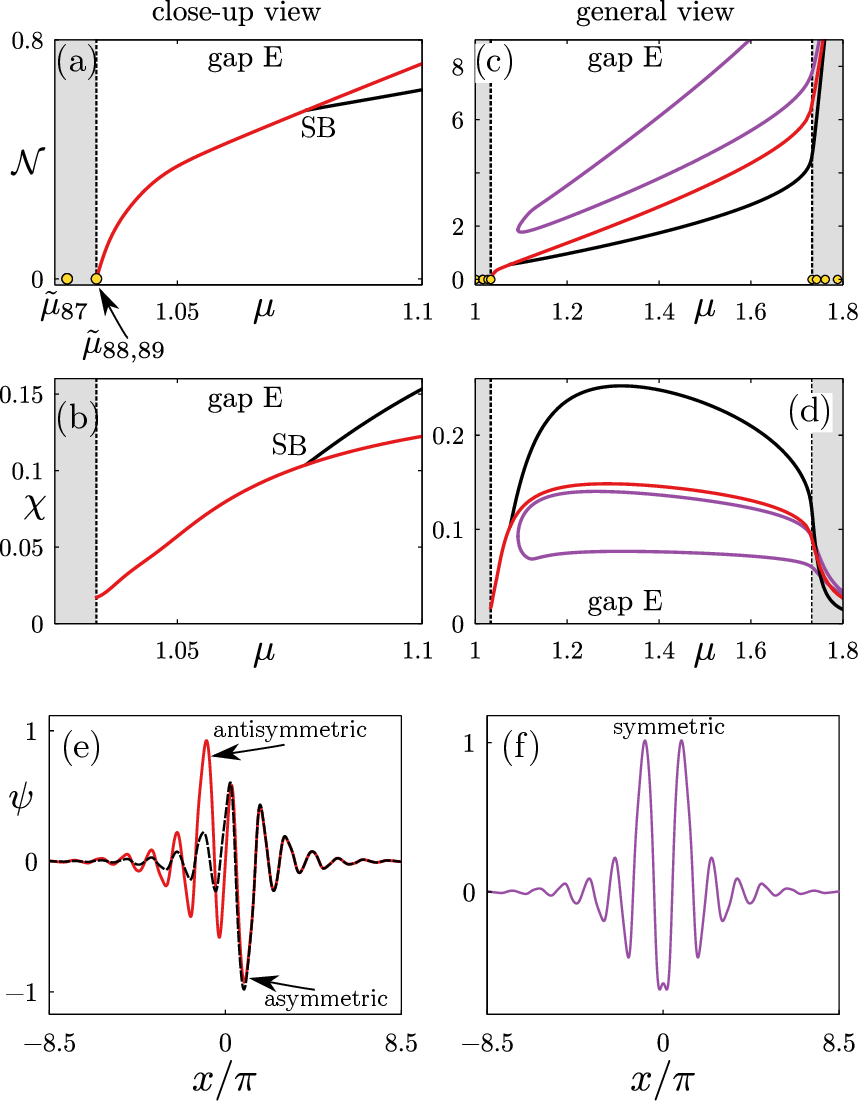}
	\end{center}
	\caption{The number of particles $\cN$ (a,c) and the IPR $\chi$  (b,d) vs. the chemical potential $\mu$ for the nonlinear modes emerging in gap~E. Panels (a,b) present a closer-up view of the band edge where the nonlinear families are formed, while   panels (c,d) provide  a smaller-scale picture.  Yellow circles in (a) indicate the chemical potentials  $\tmu_n$ of  the   noninteracting BEC, and labels `SB' in (a,b) indicate  the symmetry-breaking   bifurcation.  Panels (e,f) show the antisymmetric, asymmetric and symmetric modes  coexisting at $\mu=1.5$.  In this figure,   $p/q = 89/55$, and  $v_1 = v_2 = 0.8$.
	}
	\label{fig:gapE}
\end{figure}

\subsection{Band  with spatially extended modes}

Next, we consider the gap~E  which is situated above the mobility edge. The corresponding  band edge is formed by a pair of almost  coinciding eigenvalues associated with spatially extended eigenfunctions, see the   schematic  in Fig.~\ref{fig:scheme}E. This pair of eigenvalues produces two nonlinear families which bifurcate from the  band edge. One of these families  consists of  antisymmetric solutions which are weakly localized in the vicinity of  the bifurcation, but become more strongly localized  around  $x=0$ as the chemical potential increases towards the center of the gap. The second bifurcating family contains solutions  localized at the boundaries of the chosen spatial window.   We therefore disregard this family in the following discussion.

The family  of antisymmetric solutions  is shown in Fig.~\ref{fig:gapE}(a,b).  Since the IPRs of the linear modes above the mobility edge are much smaller than  for linear modes below the mobility edge, the slope of the bifurcating dependence  $\cN(\mu)$ is much larger, in accordance with Eq.~(\ref{eq:slope}). In fact, in this gap    dependence $\cN(\mu)$     resembles  the   root-law behavior, which  can be explained by the similarity between the found solutions and  conventional  gap solitons in purely periodic media \cite{Yang,instability}. As the bifurcated family continues along the gap, the initially extended state  becomes more localized.  In Figs.~\ref{fig:gapE}(a,b), we observe a symmetry-breaking bifurcation at a number of particles $\cN_{SB} \sim 0.5$, which is much larger than in the case of the bifurcation in Fig.~\ref{fig:gapD}, where $\cN_{SB}$ was approximately $10^{-4}$.  This can be explained by the fact that the mechanism of the symmetry-breaking bifurcation in gap~E  is essentially different  from  that encountered for localized modes in gap~D, where the bifurcation occurs as a result of the interaction between symmetric and antisymmetric modes localized at the same potential wells. For the bifurcation in gap~E,  there is no symmetric linear state involved   in the   bifurcation. However, symmetric nonlinear states can still   be found in   gap~E. These states    emerge as a result of a saddle-node bifurcation which is visible in Fig.~\ref{fig:gapE}(c,d). An example of a symmetric `gap soliton' from the subfamily with  a lower number of atoms $\cN$ is plotted in Fig.~\ref{fig:gapE}(f).

The   family of antisymmetric `gap solitons' becomes unstable after the SB-bifurcation, while the emerging asymmetric branches remain mostly stable,  up to small oscillatory instabilities. Regarding the  families  of symmetric `gap solitons', the one with a  larger number of particles is strongly unstable (with two pairs of purely real eigenvalues), while the other  one displays  weaker oscillatory instabilities.

\subsection{Cascade of bifurcations at a nonzero phase shift}
\label{sec:shift}

The results presented above were obtained for the symmetric potential (\ref{pot-irrat}) composed of two periodic lattices with a zero phase shift. In this subsection, we will consider a more general case that includes  a nonzero phase shift between the two periodic lattices. We will use  the quasiperiodic potential of the following form 
\begin{equation}
\label{pot-shift}
V(x)=v_1 \cos{(2x)}+v_2 \cos (2\varphi x + \theta ),  
\end{equation}
where  $\theta \in [0, 2\pi)$ is the phase shift between the lattices. For a generic choice of the phase shift $\theta$, this  potential  is asymmetric and does not have a global minimum or maximum.  

\begin{figure}
	\begin{center}
		\includegraphics[width=0.999\columnwidth]{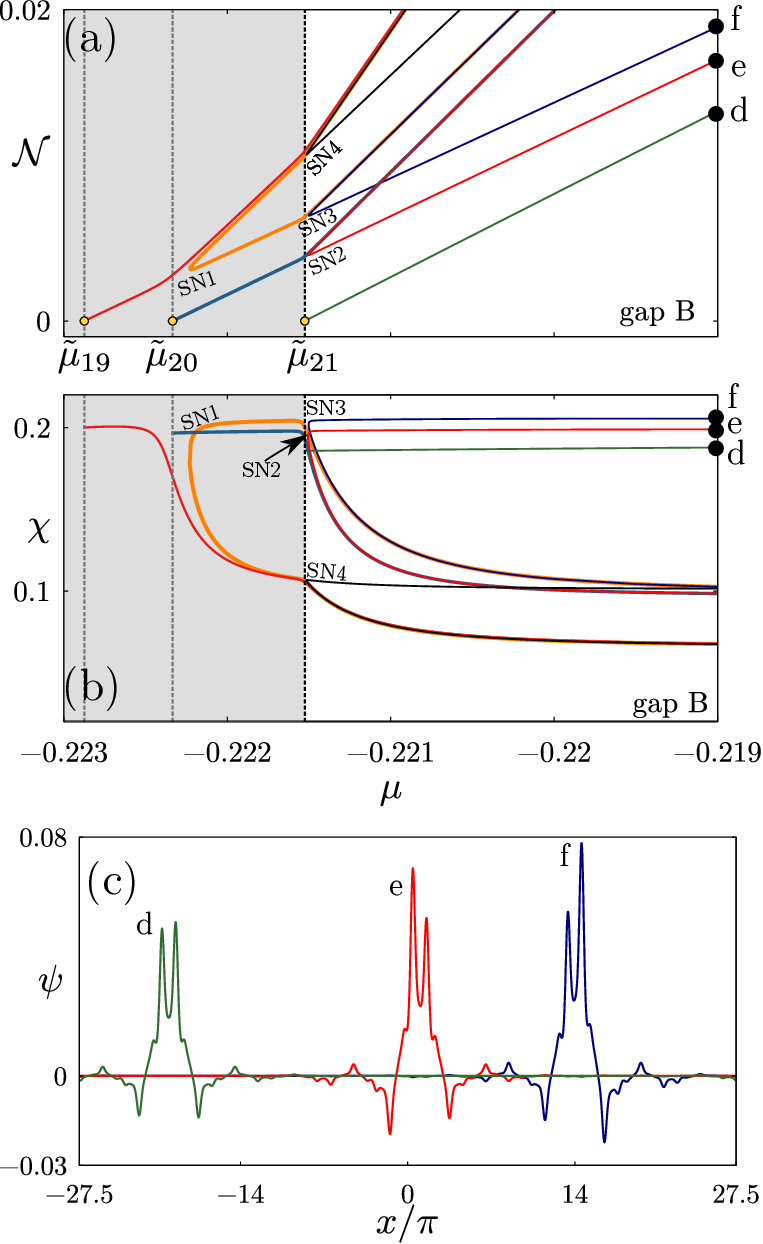}
	\end{center}
	\caption{The number of particles $\cN$ (a) and IPR $\chi$ (b) are plotted  against the  chemical potential $\mu$, for families of nonlinear modes that emerge in gap~B in  an asymmetric quasiperiodic potential described by Eq.~(\ref{pot-shift}) with a nonzero phase shift $\theta \approx 1.7 \pi$. The yellow circles in (a) show the chemical potentials $\tmu_n$ of  the   noninteracting BEC. The  labels `SN1',\ldots,`SN4' indicate    saddle-node bifurcations, and the points `d,e,f' in (a,b) correspond to three localized nonlinear modes that get through the bifurcation cascade and stay free from the hybridization with other modes.  The spatial profiles of these localized modes are shown in (c).  In this figure, the   RA with   $p/q = 89/55$ is used,   and the  lattice amplitudes are set to $v_1 = v_2 = 0.8$.}
	\label{fig:phase}
\end{figure}

On the basis of the results accumulated above,  we can expect that an effect of a  small but nonzero shift $\theta$ is  similar to   the transition from a symmetric double-well potential to a weakly asymmetric one. Respectively, symmetry-breaking bifurcations transform  to saddle-node ones \cite{TheKev}. \rev{For a generic choice of the  phase shift, the quasiperiodic potential  has no any particular symmetry, and hence the pairs of closely spaced symmetric and asymmetric eigenmodes disappear from the linear spectrum, and  the distribution of linear eigenvalues becomes more uniform within each band.} Respectively,   saddle-node bifurcations     prevail for the nonlinear modes. A fragment of the bifurcation diagram with an arbitrarily chosen phase shift $\theta$ is shown in Fig.~\ref{fig:phase}(a,b).  It displays  saddle-node bifurcations that emerge for the three rightmost eigenmodes at the of band edge near the gap~B. Each linear mode originates a nonlinear family which reaches the chemical potential corresponding to the right-nearest mode and   produces a saddle-node (SN) bifurcation, resulting in the formation of hybridized states. Moreover, secondary   bifurcations occur due to the  hybridization between linear modes and the nonlinear families that  emerge after  the `primary' bifurcations. This leads to a complex cascade of SN bifurcations, as shown in Fig. \ref{fig:phase}a and b.  At the same time, each linear mode eventually makes its way into the gap producing a localized nonlinear mode of the corresponding shape. Examples of these localized modes that get through the bifurcation cascade and emerge inside the gap are displayed in Fig.~\ref{fig:phase}(c).

\section{Discussion and conclusion}
\label{sec:concl}

In this study, we have investigated the typical mechanisms that lead to the formation of weakly nonlinear states in one-dimensional Bose-Einstein condensates confined by bichromatic quasi-periodic lattices. Transitions between linear and nonlinear regimes are well known for nonlinear modes arising from isolated eigenvalues in single- and double-well potentials, as well as for solitons forming in gaps of purely periodic potentials.  However, the spectrum of a quasiperiodic potential  with a mobility edge is organized in a more intricate way, as it contains a complex set of  eigenvalues associated with localized modes below the mobility edge and  bands of spatially extended states above the mobility edge. As a result, the full picture of the bifurcations between linear and nonlinear steady states  becomes much more involved and reach. 

The main findings of our study can be summarized as follows. Due to the presence of a mobility edge, the formation of nonlinear modes can occur in two different scenarios. In the first scenario, the underlying linear mode is already localized. In the second scenario, the linear mode is spatially extended and gradually develops localization as the chemical potential approaches the gap center. The difference between these two situations is reflected in the different dependencies of the number of atoms in the condensate on the chemical potential. In the former case, the dependence is linear. In the latter case, it is more like a square root law.
Furthermore, we have found that in a symmetric potential the birth of nonlinear modes can be associated with  either pitchfork bifurcations or saddle-node bifurcations that mimic distinctive behaviors  in  symmetric and asymmetric double-well potentials,  respectively.  A bifurcation of either type can also occur for nonlinear modes that form above the mobility edge and resemble gap solitons in periodic lattices. In the general case of an asymmetric potential, saddle-node bifurcations occur more frequently, and the formation of localized modes in the gap occurs through a series of saddle-node bifurcations.

We believe that our findings deepen the current understanding of the interaction between nonlinearity and quasiperiodicity, and reveal several important patterns that govern the bifurcation of nonlinear modes from a fractal linear spectrum. Our results were obtained in a fairly simple setup involving two superimposed laser beams, which can be implemented in experiments with cold atoms. Moreover, given the analogy between the Gross-Pitaevskii equation and the nonlinear Schrodinger equation for the paraxial propagation of light, our findings also have implications for nonlinear optics, where the fabrication of quasiperiodic photonic structures is within reach of modern techniques.

\begin{acknowledgments}
The research    is financially supported by Russian Science Foundation, grant no. 23-11-00009, https://rscf.ru/project/23-11-00009/.
 
\end{acknowledgments}


\begin{thebibliography}{}
	
	
	\bibitem{Lye} J. E. Lye, L. Fallani, C. Fort, V. Guarrera, M. Modugno,  D. S. Wiersma, and M. Inguscio, Effect of interactions on the localization of a Bose-Einstein condensate in a quasiperiodic lattice, Phys. Rev. A {\bf  75}, 061603(R) (2007).
	
	\bibitem{Roati} G. Roati, C. D’Errico, L.  Fallani, M. Fattori, C.  Fort, M.  Zaccanti, G.  Modugno, M.  Modugno, and   M.   Inguscio, Anderson localization of a non-interacting
	Bose–Einstein condensate, Nature 453,  {\bf  895}  (2008).
	
	\bibitem{Modugno10} G. Modugno, Anderson localization in Bose–Einstein 	condensates, Rep. Prog. Phys. {\bf  73},  102401  (2010).
	
	\bibitem{Reeves}   J. B. Reeves, B. Gadway, T. Bergeman, I. Danshita,
	and D. Schneble, Superfluid Bloch dynamics in an incommensurate optical lattice, New J. Phys. {\bf  16}, 065011
	(2014).
	
\bibitem{Luschen}  H. P. L\"uschen, S. Scherg, T. Kohlert, M. Schreiber, P. Bordia,
	X. Li, S. Das Sarma, and I. Bloch, Single-particle mobility edge
	in a one-dimensional quasiperiodic optical lattice, Phys. Rev.
	Lett. {\bf  120}, 160404 (2018).
		
	
	
	\bibitem{Tanese} D. Tanese, E. Gurevich, F. Baboux, T. Jacqmin, A. Lema\^{\i}tre, E. Galopin, I. Sagnes, A. Amo, J. Bloch, and E. Akkermans, Fractal Energy Spectrum of a Polariton Gas in a Fibonacci Quasiperiodic Potential, Phys. Rev. Lett. {\bf 112}, 146404  (2014).
	
	
	\bibitem{Goblot} V. Goblot, A. \v{S}trkalj, N. Pernet, J. L. Lado, C. Dorow, A. Lema\^{\i}tre, L. Le Gratiet, A. Harouri, I. Sagnes, S. Ravets, A. Amo, J. Bloch, and  O. Zilberberg, Emergence of criticality through a cascade of
	delocalization transitions in quasiperiodic chains, Nature Phys. {\bf 16},  832--836 (2020).
	
		
	
	\bibitem{Lahini} Y. Lahini, R. Pugatch,  F. Pozzi,  M. Sorel, R. Morandotti,  N. Davidson,  and Y. Silberberg, Observation of a Localization Transition in Quasiperiodic Photonic Lattices, Phys. Rev. Lett.   {\bf  103}, 013901 (2009).
	
	\bibitem{Arie2010} 	 A. Arie and N. Voloch, Periodic, quasi-periodic, and random quadratic nonlinear photonic crystals, Laser Photon. Rev. {\bf  4}, 355 (2010).
	
	
	
		\bibitem{Bellingeri17}  M. Bellingeri, A. Chiasera, I. Kriegel, and F. Scotognella, Optical properties of periodic, quasi-periodic, and disordered onedimensional photonic structures, Opt. Mater.
{\bf  72}, 403 (2017).

\bibitem{Valy13}
Z. Valy Vardeny, A.  Nahata, and A.  Agrawal, Optics of photonic quasicrystals, Nat. Photonics {\bf  7}, 177
(2013).
	
	
	\bibitem{Wang2022} Y.  Wang,  J.-H. Zhang, Y. Li, J.  Wu, W.  Liu,  F.   Mei, Y. Hu, 	L.  Xiao,  J.   Ma,  C.  Chin,  and S.  Jia, Observation of Interaction-Induced Mobility Edge in an Atomic Aubry-Andr\'e Wire, Phys. Rev. Lett. {\bf  129}, 103401 (2022).
	
		\bibitem{Segev}  M. Segev, Y. Silberberg, and  D.   N. Christodoulides, Anderson localization of light, Nature Photonics {\bf 7}, 197--204 (2013).
		
		\bibitem{Carretero} R. Carretero-Gonz\'alez,  D. J. Frantzeskakis, and P. G. Kevrekidis, Nonlinear waves in Bose–Einstein condensates:
		physical relevance and mathematical techniques, Nonlinearity {\bf 21},  R139  (2008).
		
		\bibitem{Longhi} S. Longhi, Quantum-optical analogies using photonic structures, Laser \& Photon. Rev. {\bf  3}, 243  (2009).
	
	\bibitem{Sakaguchi2006} H. Sakaguchi and B. A. Malomed, Gap solitons in quasiperiodic optical lattices, Phys. Rev. E \textbf{74},
	026601 (2006).
	
	\bibitem{Huang19} 	C. Huang, C. Li, H. Deng, and L. Dong, Gap Solitons in Fractional Dimensions With a Quasi-Periodic Lattice,  Ann. Phys. {\bf  531},  1900056
	(2019).
	
	
	\bibitem{Kominis2008} Y. Kominis and K. Hizanidis, Power dependent soliton location and
	stability in complex photonic structures, Opt. Express {\bf 16}, 12124 (2008).
	
	\bibitem{Baizakov}
	B. B. Baizakov, B. A. Malomed, and M. Salerno, in \textit{ Nonlinear
	Waves: Classical and Quantum Aspects}, edited by F. Kh. 	Abdullaev and V. V. Konotop (Kluwer Academic, Dordrecht,
	2004), p. 61.
	
	\bibitem{Burlak07} 	G. Burlak and A. Klimov,  The solitons redistribution in Bose--Einstein condensate
	in quasiperiodic optical lattice, Phys. Lett. A {\bf  369}, 510 (2007).
	
	\bibitem{Burlak12} G. Burlak and B.  A. Malomed, Matter-wave solitons with the minimum number of particles in two-dimensional
	quasiperiodic potentials, Phys. Rev. E {\bf  85}, 057601 (2012).
	
	
	\bibitem{Xie03} P. Xie, Z.-Q. Zhang, and X. Zhang, Gap solitons and soliton trains in finite-sized two-dimensional periodic
	and quasiperiodic photonic crystals,  Phys. Rev. E {\bf  67}, 026607 (2003).
	
	\bibitem{Ablowitz06} M. J. Ablowitz, B. Ilan, E. Schonbrun,  and R. Piestun, Solitons in two-dimensional lattices possessing defects, dislocations, and quasicrystal structures, Phys. Rev. E  {\bf 74}, 035601(R) (2006).
	
	\bibitem{Ablowitz10} M. J. Ablowitz, N. Antar, \.{I}. Bakirta\c{s}, and B. Ilan, Band-gap boundaries and fundamental solitons in complex two-dimensional nonlinear lattices, Phys. Rev. A  {\bf 81}, 033834 (2010).
	
	\bibitem{Ablowitz12} M. J. Ablowitz, N. Antar, \.{I}. Bakirta\c{s}, and B. Ilan, Vortex and dipole solitons in complex two-dimensional nonlinear lattices, Phys. Rev. A  {\bf 86}, 033804 (2012).
	
	\bibitem{Huang16} 	C. Huang, F. Ye, X. Chen, Y. V. Kartashov, V. V. Konotop, and L. Torner, Localization-delocalization wavepacket transition in Pythagorean aperiodic potentials,
	Sci. Rep. {\bf  6}, 32546 (2016).
	
	\bibitem{Huang21} C. Huang, L. Dong, H. Deng, X.  Zhang, P.   Gao, Fundamental and vortex gap solitons in quasiperiodic photonic lattices, Opt. Lett. {\bf 46}, 5691 (2021). 
	
	\bibitem{Kartashov21} 	Y. V. Kartashov, F. Ye,  V. V. Konotop, and Ll.  Torner,  Multifrequency Solitons in Commensurate-Incommensurate Photonic Moir\'e Lattices, Phys. Rev. Lett. {\bf 127}, 163902 (2021).
	
	\bibitem{Bagchi2012} M. Bag\v{c}hi, Soliton dynamics in quadratic nonlinear media
	with two-dimensional Pythagorean aperiodic
	lattices, JOSA B {\bf 38}, 1276 (2021).
	
	\bibitem{Kartashov23} S. K. Ivanov, V. V. Konotop, and Y. V. Kartashov, 	
	Vortex solitons in moir\'e optical lattices, 
	Opt Lett. {\bf 48}, 3797--3800 (2023). 
	
	
	\bibitem{Liu2024} X. Liu,  J. Zeng, Two-dimensional localized modes in nonlinear 	systems with linear nonlocality and moir\'e lattices,  Front. Phys. {\bf  19}, 42201 (2024).
	
	\bibitem{Wang2020}  P. Wang, Y. Zheng, X. Chen, C. Huang, Y. V. Kartashov, Ll.   Torner, V.   V. Konotop, and F.   Ye, Localization and delocalization of light in photonic moiré lattices, Nature  {\bf  577}, pages42–46 (2020).
	
	\bibitem{Fu2020} Q. Fu, P. Wang, C. Huang, Y. V. Kartashov, Ll. Torner, V. V. Konotop, and   F. Ye, Optical soliton formation controlled by angle twisting in photonic moiré lattices, Nat. Photonics {\bf  14},  663  (2020).
	
	
	\bibitem{Freedman06} 	B. Freedman, G. Bartal, M.   Segev, R.   Lifshitz, D. N. Christodoulides,   J.   W. Fleischer, Wave and defect dynamics in nonlinear photonic quasicrystals, Nature   {\bf  440}, 1166–1169 (2006).
	
	
	\bibitem{Johan1} M. Johansson and R. Riklund, Solitonlike states in a one-dimensional nonlinear Schr\"odinger equation with a deterministic aperiodic potential, Phys. Rev. B {\bf 49}, 6587 (1994).
	
	\bibitem{Johan2} B. Lindquist, M. Johansson and R. Riklund, Soliton dynamics and interaction in a deterministic aperiodic nonlinear lattice, Phys. Rev. B {\bf 50}, 9860 (1994).
	
	\bibitem{Sukho06} A. A. Sukhorukov, Enhanced Soliton Transport in Quasiperiodic Lattices with Introduced Aperiodicity, Phys. Rev. Lett. {\bf  96}, 113902 (2006).
	
	
	\bibitem{Marti10} A. J. Mart\'{\i}nez and M. I. Molina, Surface solitons in quasiperiodic nonlinear photonic lattices, Phys. Rev. A  {\bf  85}, 013807 (2012).
	
	\bibitem{Grig10} V. V. Grigoriev, F. Biancalana, Bistability and stationary gap solitons in quasiperiodic
	photonic crystals based on Thue–Morse sequence, Photonics and Nanostructures – Fundamentals and Applications {\bf  8}, 285  (2010).
	
	\bibitem{Huang21OE} C.  Huang, Z.   Lin, L.   Dong, C.   Li, P.  Gao, and W.   Su, Fundamental and multipole solitons in
	amplitude-modulated Fibonacci lattices, Opt. Express {\bf 29},  35327 (2021).
	
	\bibitem{Su22} 	W. Su,  Z.  Lin, C. Li, C.  Huang, Nonlinear localized gap modes in width-modulated Fibonacci lattices, Results in Physics, {\bf 40},    105877 (2022).
	
	\bibitem{Takahashi12} 	 	M. Takahashi, H. Katsura, M. Kohmoto, and T. Koma,	Multifractals competing with solitons on Fibonacci optical lattices, New J. Phys.  {\bf  14}, 113012 (2012).
	
	

	
	\bibitem{Kevrekidis2006} P. G. Kevrekidis, V. V. Konotop, A. Rodrigues, and D. J. Frantzeskakis, Dynamic generation of matter solitons from linear
	states via time-dependent scattering lengths, J. Phys. B: At. Mol. Opt. Phys. {\bf 38},  1173  (2005).
	
	
	\bibitem{BraKon} V. A. Brazhnyi and V. V. Konotop, Theory of nonlinear matter waves in optical lattices. Mod. Phys. Lett. B {\bf 18}, 627 (2004).
	
	\bibitem{Deissler10} 	B. Deissler, M. Zaccanti, G. Roati, C. D’Errico, M. Fattori, M. Modugno, G. Modugno, and M. Inguscio, Delocalization of a disordered bosonic system by repulsive interactions, Nat. Phys. {\bf  6},  354  (2010).
	
	\bibitem{Zilber2021} O. Zilberberg, Topology in quasicrystals [Invited], Opt. Mater.
	Express {\bf 11}, 1143 (2021).
	
	
	\bibitem{Diener} R. B. Diener, G. A. Georgakis, J. Zhong, M. Raizen, and Q. Niu
	Transition between extended and localized states in a one-dimensional incommensurate optical lattice, Phys. Rev. A {\bf 64}, 033416 (2001)  
	
	
	
	\bibitem{Modugno} M. Modugno, Exponential localization in one-dimensional quasi-periodic optical lattices, New J. Phys. {\bf 11}, 033023 (2009).
	
	\bibitem{LiLiSarma2017} X. Li, X. Li, and S. Das Sarma, Mobility edges in one-dimensional bichromatic incommensurate potentials, Phys.
	Rev. B  {\bf  96}, 085119 (2017).
	
	
	\bibitem{Prates22} H. C. Prates,  D. A. Zezyulin,  and V. V. Konotop, Bose-Einstein condensates in quasiperiodic lattices: Bosonic Josephson junction, self-trapping, and multimode dynamics, Phys. Rev. Research {\bf  4}, 033219 (2022).
	
	
	\bibitem{Yang} J. Yang, \textit{Nonlinear waves in integrable and nonintegrable systems}, Philadelphia: SIAM, 2010.
	
	
	\bibitem{Boers2007} D. J. Boers, B. Goedeke, D. Hinrichs, and M. Holthaus, Mobility edges in bichromatic optical lattices, Phys. Rev. A \textbf{75}, 063404 (2007).
	
	\bibitem{Roux2008} G. Roux, T. Barthel, I. P. McCulloch,  C. Kollath, U. Schollw\"ock, and T. Giamarchi, Quasiperiodic Bose-Hubbard model and localization in one-dimensional cold atomic gases, Phys. Rev.  A {\bf  78}, 023628 (2008).
	
	\bibitem{Larcher2009}  M. Larcher,  F. Dalfovo,  and M. Modugno, Effects of interaction on the diffusion of atomic matter waves 	in one-dimensional quasiperiodic potentials, Phys. Rev. A {\bf  80}, 053606 (2009).
	
	
	\bibitem{Biddle} J. Biddle, B. Wang, D. J. Priour, and S. Das Sarma, Localization in one-dimensional incommensurate lattices beyond the Aubry-Andr\'e model. Phys. Rev. A {\bf 80}, 021603(R) (2009).
	
		
	\bibitem{Adhikari2009} S. K. Adhikari and L. Salasnich, Localization of a Bose-Einstein condensate in a bichromatic optical lattice, Phys. Rev.
	A {\bf  80}, 023606 (2009).
	
	\bibitem{BidSar} J. Biddle and S. Das Sarma, Predicted Mobility Edges in One-Dimensional Incommensurate Optical Lattices:
	An Exactly Solvable Model of Anderson Localization, Phys. Rev. Lett. {\bf 104}, 070601 (2010).
	
	
	\bibitem{Zhou13}  L. Zhou,  H. Pu, and W.  Zhang, Anderson localization of cold atomic gases with effective spin-orbit interaction      in a quasiperiodic optical lattice, Phys. Rev. A  {\bf  87}, 023625 (2013).
	
	
	\bibitem{Cheng14}  Y.   Cheng,  G.  Tang, and S. K. Adhikari, Localization of a spin-orbit-coupled Bose-Einstein condensate in a bichromatic optical lattice, Phys. Rev. A {\bf   89}, 063602 (2014).
	
	\bibitem{Li16}    C.  Li, F.   Ye, Y.   V. Kartashov, V.   V. Konotop, and   X.   Chen, Localization-delocalization transition in spin-orbit-coupled Bose-Einstein condensate, Sci. Rep. {\bf  6},  31700 (2016).
	
	
	\bibitem{Yao2019} H. Yao, H. Khoudli, L. Bresque, and L. Sanchez-Palencia, Critical Behavior and Fractality in Shallow One-Dimensional Quasiperiodic Potentials, Phys. Rev. Lett. \textbf{123}, 070405 (2019).
	
	\bibitem{Konotop2024} V. V. Konotop, Dimers and discrete breathers in Bose-Einstein condensates in a quasi-periodic potential, Phys. Rev. Research {\bf   6}, 033113 (2024).
	
	\bibitem{Zez08} D. A. Zezyulin, G. L. Alfimov, V. V. Konotop, and V. M. P\'{e}rez--Garc\'{\i}a,  Stability of excited states of a Bose-Einstein condensate in an anharmonic trap, Phys. Rev. A {\bf 78},    013606 (2008).
	
	
	\bibitem{SBB1} R. D'Agosta and C. Presilla, States without a linear counterpart
	in Bose-Einstein condensates, Phys. Rev. A {\bf 65}, 043609 (2002).
	
	\bibitem{SBB2} R. K. Jackson and M. I. Weinstein, Geometric analysis of bifurcation and symmetry breaking in a Gross-Pitaevskii equation, J. Stat. Phys. {\bf 116}, 881 (2004).
	
	\bibitem{SBB3} P. G. Kevrekidis, Z. Chen, B. A. Malomed, D. J. Frantzeskakis,
	and M. I. Weinstein, Spontaneous symmetry breaking in photonic lattices: Theory and experiment, Phys. Lett. A {\bf  340}, 275 (2005).
	
		
	\bibitem{TheKev} G. Theocharis, P. G. Kevrekidis, D. J. Frantzeskakis, and P. Schmelcher, Symmetry breaking in symmetric and asymmetric double-well potentials, Phys. Rev. E {\bf 74}, 056608 (2006).
			
	\bibitem{SBB5}  G. L. Alfimov and D. A. Zezyulin, Nonlinear modes for the
	Gross–Pitaevskii equation --- a demonstrative computation approach, Nonlinearity {\bf 20}, 2075 (2007).
	
	
	\bibitem{SBB6} M. Matuszewski, B. A. Malomed, and M. Trippenbach, Spontaneous symmetry breaking of solitons trapped in a double-channel potential, Phys. Rev. A {\bf  75}, 063621 (2007)
	
	
	\bibitem{SBB7}  E. W. Kirr, P. G. Kevrekidis, E. Shlizerman, and
	M. I. Weinstein, Symmetry-breaking bifurcation in nonlinear
	Schrödinger/Gross–Pitaevskii equations, SIAM J. Math. Anal.
	{\bf  40}, 566 (2008).
	
	\bibitem{SBB8} A. Sacchetti, Universal Critical Power for Nonlinear
	Schrödinger Equations with a Symmetric Double Well
	Potential, Phys. Rev. Lett. {\bf  103}, 194101 (2009).
		
	\bibitem{SBB9} E. Shamriz, N. Dror, and B. A. Malomed, Spontaneous symmetry breaking in a split potential box, Phys. Rev. E {\bf  94}, 022211 (2016).
	
\bibitem{instability}
P. J. Y. Louis, E. A. Ostrovskaya, C. M. Savage, Yu. S. Kivshar, Bose–Einstein condensates in optical lattices: Band-gap structure and solitons, Phys. Rev. A {\bf  67}, 013602  (2003); N. K. Efremidis, D. N. Christodoulides, Lattice solitons in Bose-Einstein condensates, Phys. Rev. A \textbf{67}, 063608 (2003); D. E. Pelinovsky, A. A. Sukhorukov, Yu. S. Kivshar, Bifurcations and stability of gap solitons in periodic potentials, Phys. Rev. E {\bf 70}, 036618  (2004);
P. P. Kizin, D. A. Zezyulin,  G. L. Alfimov, Oscillatory instabilities of gap solitons in a repulsive Bose–Einstein condensate, Physica D {\bf 337}, 58 (2016).

\bibitem{Kapitula2006} T. Kapitula, P. G. Kevrekidis, and Z. Chen, Three is a crowd: Solitary waves in photorefractive media with three potential
wells, SIAM J. Appl. Dyn. Syst. {\bf 5}, 598 (2006).


\bibitem{Sacchetti2012}  A. Sacchetti, Nonlinear Schrödinger equations with multiple-well potential, Physica D  {\bf 241}, 1815 (2012).


\bibitem{Goodman2017} R. H. Goodman, Bifurcations of relative periodic orbits in NLS/GP with a triple-well
potential, Physica D {\bf 359}, 39 (2017). 



\bibitem{composition} Yo. Zhang, Zh. Liang and B. Wu, Gap solitons and Bloch waves in nonlinear periodic systems,
Phys. Rev. A {\bf  80},  063815  (2009).

\bibitem{Akylas2012} T. R. Akylas, G. Hwang, and J. Yang, From non-local gap solitary waves to bound
states in periodic media, Proc. R. Soc. A {\bf  468},  116 (2012).


\end{thebibliography}
\end{document}